\documentclass[aps,twocolumn,footinbib,longbibliography]{revtex4-1}
\usepackage{amsmath,amssymb}
\usepackage{bm}
\usepackage{epsfig}
\usepackage{natbib}
\usepackage{hyperref}
\usepackage{color}
\usepackage{graphicx}

\hypersetup{pdfstartview={FitH},pdfpagemode={UseNone},
            colorlinks,linkcolor=red, citecolor=blue, urlcolor=blue,
            bookmarks=true, bookmarksopen=true, pdfnewwindow=true}

\usepackage{dcolumn}

\newcommand{\rmi}{{\rm i}}

\newcommand {\e}{{\rm e}}

\begin{document}


\title{Optical activity in chiral stacks of 2D semiconductors\\
}

\author{Alexander V. Poshakinskiy}
\email{poshakinskiy@mail.ioffe.ru}
\author{Dmitrii R. Kazanov}
\author{Tatiana V. Shubina}
\author{Sergey A. Tarasenko}
\affiliation{Ioffe Institute, 194021 St.~Petersburg, Russia}

\date{\today}

\begin{abstract}
We show that the stacks of two-dimensional semiconductor crystals with the chiral packing exhibit optical activity and circular dichroism. We develop a microscopic theory of these phenomena in the spectral range of exciton transitions which takes into account the spin-dependent hopping of excitons between the layers in the stack and the interlayer coupling of excitons via electromagnetic field. For the stacks of realistic two-dimensional semiconductors such as transition metal dichalcogenides, we calculate the rotation and ellipticity angles of radiation transmitted through such structures. The angles are resonantly enhanced at the frequencies of both bright and dark exciton modes in the stack. We also study the photoluminescence of chiral stacks and show that it is circularly polarized.
\end{abstract}

\pacs{Valid PACS appear here}
                             
           
\maketitle


\section{Introduction}

Optical activity, the ability of certain media to rotate the plane of light polarization in the absence of a magnetic field, is a remarkable manifestation of polarization dependent interaction between light and matter~\cite{Goldstein_Book}. It stems from the different strengths of coupling to right-handed and left-handed circularly polarized light leading to circular birefringence and circular dichroism~\cite{Kaminsky2000}. From the pioneering experiment
by J.C. Bose in 1898 with left- and right-twisted jute elements~\cite{Bose1898}, the optical rotation is commonly associated with chirality. Optically active media attract much attention because they allow one to manipulate the polarization state of light. 
This is particularly important for superresolution imaging and the development of broadband optical components, such as filters and sensors with high polarization suppression ratio~\cite{Wang2017,Chadha2014}. Moreover, the photon polarization is currently considered 
as the degree of freedom to code and carry quantum information~\cite{Bennett1992}. 


Optical activity, also referred to as gyrotropy, can be of either intrinsic or extrinsic nature. The former occurs in macroscopically homogeneous systems with chiral molecule components or crystals of the gyrotropic symmetry classes~\cite{Kizel1975}.
A canonical example of optically active biological media is the tartaric acid, which was discovered by L. Pasteur and explained by the predominance of one of the two possible stereoisomers of the acid molecules~\cite{Mason_Book}. Tellurium is an example of gyrotropic crystals~\cite{Nomura1960}. This elemental semiconductor can exist in two enantiomorphic forms with the atoms, constituting the crystal lattice, bound into left-handed or right-handed chains. An example of gyrotropic but not chiral bulk crystals is silver gallium sulphide~\cite{Hobden1967}.  
Microscopically, optical activity originates from the spatial dispersion of susceptibility~\cite{Agranovich_Book}. It is increased
in the vicinity of exciton resonances, as was observed in bulk wurtzite crystals~\cite{lvchenko1979} and II-VI quantum well structures~\cite{Kotova2016}. The term ``extrinsic optical activity'' coined recently is primarily used for metamaterials, metasurfaces, and photonic crystals composed of the arrays of chiral or achiral elements such as gammadions~\cite{Rogacheva2006,Lobanov2015,Demenev2016}, G-shape nanostructures~\cite{Mamonov2014}, and split-ring resonators~\cite{Plum2009}. The efficiency of polarization conversion can be enhanced by plasmonic resonances in metal nanostructures~\cite{Minovich2015,Collins2017}.

Apart from the above two distinct types of gyrotropic structures, we highlight the opportunity of an intermediate case of artificial materials with chiral stacking of atomic layers. 
Previously, they would be hypothetical.
The emerging technology of van der Waals structures made of two-dimensional (2D) crystals of atomic thickness~\cite{Geim2013,lotsch2015} enables the formation of such materials. Recently, it has been experimentally demonstrated that a pair of graphene layers stacked with a twist exhibits the optical activity~\cite{Kim2016}. The optical rotation angle normalized to the sample thickness was found to be several orders of magnitude larger than that in natural materials. The 2D crystals beyond the graphene, such as transition metal dichalcogenides (TMDC) MoS$_2$, WS$_2$, etc., are of particular interest for application in optics. TMDC layers have optical gaps and demonstrate strong light-matter coupling in the spectral range of exciton transitions~\cite{Mak2010,Mak2016,Cadiz2017,Jun2017}. Pronounced exciton resonances have been observed in the reflectance spectra of TMDC layers and thin films~\cite{Arora2015,Maciej2017}. The properties of twisted stacks 
including the interlayer coupling strength are also being widely studied~\cite{Liu2014, Huang2014, Zande2014,Yan2015,Plechinger2015,Xia20171}. At the same time, the optical polarization effects related to chirality slipped away from the focus of previous research.

\begin{figure}[t]
  \includegraphics[width=.99\columnwidth]{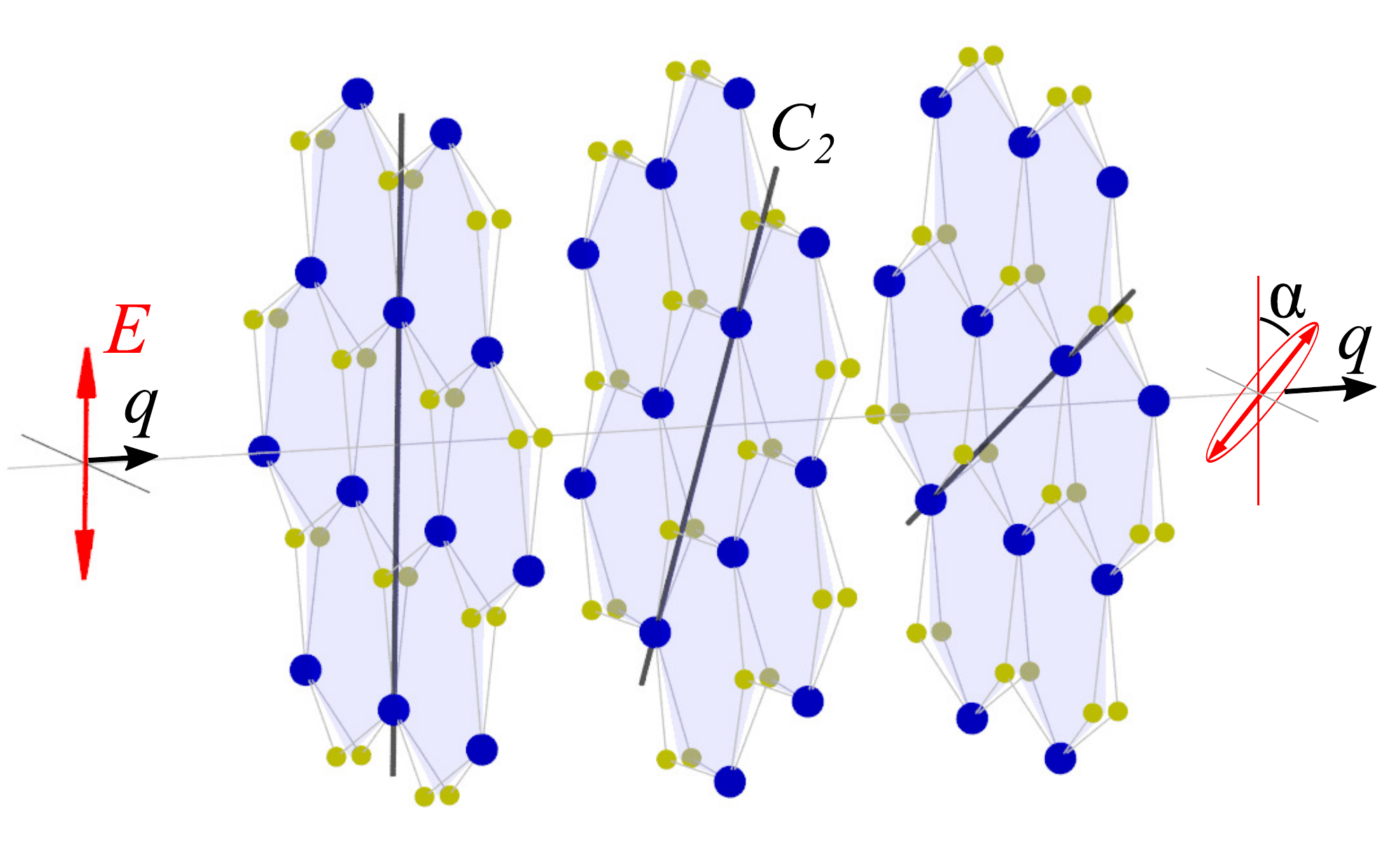}
\caption{A sketch of a chiral stack of TMDC layers. Linearly polarized light passing through the
chiral stack rotates its polarization plane and acquires ellipticity.}\label{fig:sketch}
\end{figure}

In this paper, we study the optical properties of chiral stacks of 2D gapped crystals. We show that properly arranged stacks (see Fig.~\ref{fig:sketch}) exhibit
pronounced optical activity while individual layers do not. We develop a microscopic theory of the optical activity and the accompanying phenomenon of circular dichroism. It is also predicted that the photoluminescence of chiral stacks is circularly polarized.

The paper is organized as follows. In Sec.~\ref{Sec2}, we develop a theoretical approach to describe the exciton states in the chiral stacks of TMDC layers and the polarization optical properties of such stacks. We start by considering chiral bilayers and then generalize the model to multi-layer stacks. In Sec.~\ref{Sec3}, we describe the spectral behavior of light transmission, optical activity, and circular dichroism for the chiral bilayers, multi-layers, and thick TMDC stacks. In Sec.~\ref{Sec4}, we discuss the circularly polarized photoluminescence of chiral stacks. Section~\ref{Sec5} summarizes the results of the paper.

\section{Model}\label{Sec2}


The spatial symmetry of a single TMDC layer is described by the $D_{3h}$ point group. The layers are optically inactive and demonstrate 
isotropic linear response at the normal incident of radiation. We consider the optical response associated with bright direct excitons with the spin projections $s = \pm 1$ along the normal to the layer which are excited by the $\sigma^{\pm}$ circularly polarized light, respectively. The excitons are formed by electrons and holes located at the $K$ and $K'$ valleys of the two-dimensional Brillouin zone. The exciton states with the spin projections $s = \pm 1$ at zero in-plane wave vector are degenerate in energy.

Arranging two TMDC layers into a twisted stack reduces the point-group symmetry of the system to $D_3$. The exciton states in such a bilayer are described by an effective Hamiltonian $H_{ns,n's'}$, where $n,n'=1,2$ stand for the monolayer index and $s,s'=\pm 1$ stand for the spin index. We note that, in multi-layer systems, there are also excitations consisting of electrons and holes localized in different layers~\cite{Arora2017,Nagler2017}. Those interlayer excitons are far separated in energy from the intralayer excitons because of the reduced Coulomb interaction and are not considered here. 

The $D_3$ point group of the bilayer imposes restrictions on the form of the effective exciton Hamiltonian. The presence of the three-fold rotation axis along the bilayer normal eliminates the matrix elements of the Hamiltonian with $s\neq s'$, so that $H_{ns,ns'} = H_{n,n'}^{(s)} \delta_{s,s'}$. The two-fold rotation axes lying in the bilayer plane impose the constraints $H_{1,1}^{(s)}=H_{2,2}^{(-s)}$ and $H_{1,2}^{(s)}=H_{2,1}^{(-s)}$. Finally, the time-reversal symmetry requires $H_{n,n'}^{(s)}=H_{n',n}^{(-s)}$. Combining all the above symmetry constraints together with the requirement that the Hamiltonian is Hermitian we arrive to the most general form of the exiton Hamiltonian in a chiral bilayer
\begin{align}\label{ham:bi}
H^{(s)} = \left( \begin{array}{cc} 
\omega_x & -t \e^{\rmi s \phi} \\
 -t \e^{-\rmi s \phi} & \omega_x 
\end{array}\right) \,
\end{align} 
with real parameters $\omega_x$, $t$, and $\phi$. Here, $\omega_x$ is the exciton frequency in an isolated layer, $t$ describes the interlayer hopping of excitons due to, e.g., tunneling or F\"orster excitation transfer, and $\phi$ is the phase acquired at the interlayer hopping due to the chirality of the bilayer. The Planck constant is set to unity. The dependence of the phase $\phi$ on the parameters of particular TMDC bilayers
can be obtained from microscopic calculations which are beyond the scope of the present paper. 
Instead, we focus below on the general optical properties of chiral bilayers and multi-layers. 
We only note that the phase $\phi$ vanishes in achiral stacks which are realized at the twist angle 0, $\pi/3$, $2\pi/3$, etc.

The model above can be readily generalized to the chiral stack of $N$ layers. Assuming the hopping of excitons between the 
nearest-neighbor layers only, we shall use the exciton Hamiltonian with the matrix elements
\begin{align}\label{ham:N}
H_{n,n'}^{(s)} = \omega_x \delta_{n,n'} - t \e^{\rmi s \phi} \delta_{n+1,n'} - t \e^{-\rmi s \phi} \delta_{n-1,n'}
\end{align}
with $n,n' = 1 \ldots N$. 

\begin{figure*}[t]
\includegraphics[width=.99\textwidth]{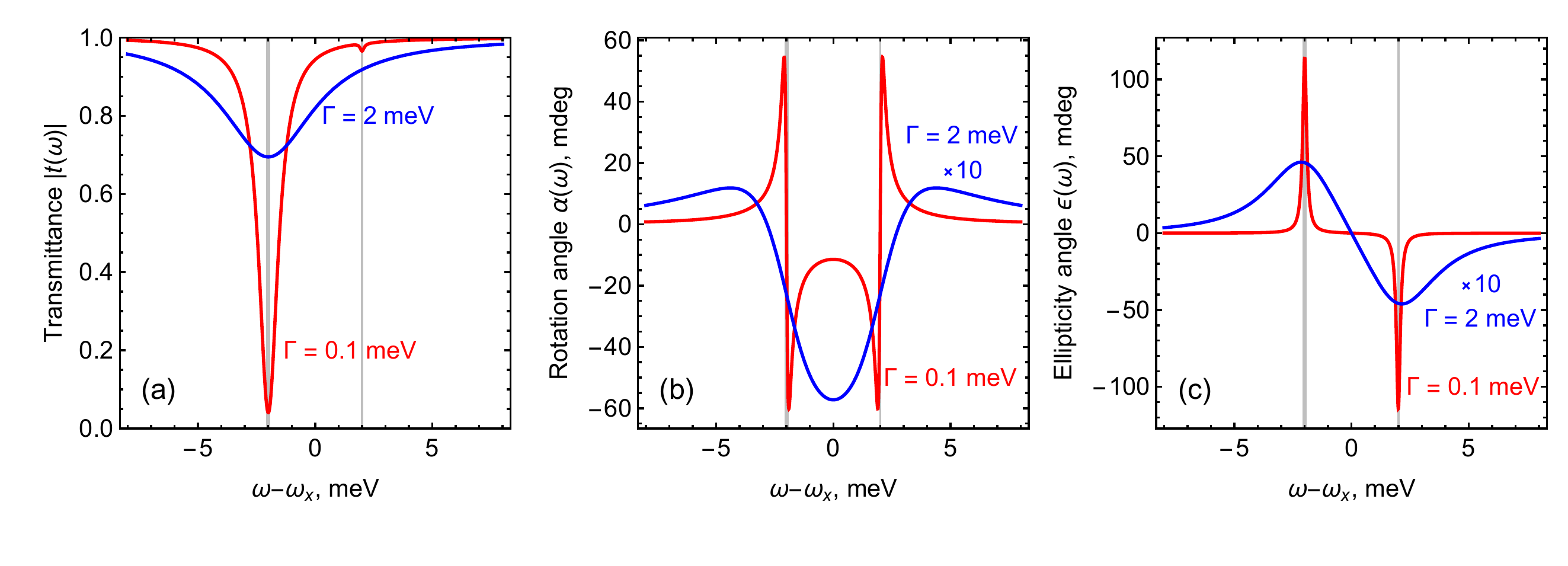}
\caption{(a) Transmittance $|t(\omega)|^2$, (b) polarization rotation angle $\alpha(\omega)$, and (c) ellipticity angle $\epsilon(\omega)$ for a chiral bilayer ($N=2$). Vertical lines indicate the frequencies of the symmetric (thick line) and anti-symmetric (thin line) exciton modes in the bilayer. The dependences are calculated for the parameters $\Gamma_0=0.3$\,meV~\cite{Robert2016b}, $t=2$\,meV, $qd=0.01$, $\phi=0.1$, and the non-radiative exciton decay rates $\Gamma=0.1$\,meV (red curves) and 2\,meV (blue curves).}\label{fig:bilayer}
\end{figure*}

Now we consider the interaction of excitons with the electromagnetic field. Due to high symmetry of an individual layer, the strength of exciton-photon coupling in a layer is described by a single parameter $\Gamma_0$ which determines the exciton radiative decay rate. For the incident light propagating along the $z$ axis (Fig.~\ref{fig:sketch}), the exciton polarizations in the layers $P_n^{(s)}$ are determined by the equation set (see Ref.~\cite{Ivchenko2005})
\begin{align}\label{sys:bi}
(\omega + \rmi\Gamma) P_n^{(s)} - \sum_{n'} (H_{n,n'}^{(s)} - \rmi \Gamma_0 \e^{\rmi q d|n-{n'}|} ) P_{n'}^{(s)} \nonumber\\
= - \frac{\Gamma_0}{2 \pi q} \, {\cal E}_{{\rm 0}}^{(s)} \e^{\rmi q n d} \,,
\end{align}
where ${\cal E}_{{\rm 0}}^{(s)}$ is the electric field amplitude of the incident radiation with a certain helicity, $\omega = c q$ and $q$ are the photon frequency and wave vector, respectively, $n d$ is the position of the $n$-th layer, and $d$ is the interlayer distance. The last term in the left-hand side of Eq.~\eqref{sys:bi} describes both the radiative decay of excitons (at $n=n'$) and
the radiative coupling of excitons in different layers (at $n \neq n'$). Equation~\eqref{sys:bi} also takes into account the non-radiative decay of excitons with the rate $\Gamma$~\cite{Ivchenko1994,Poshakinskiy2012,Kazanov2017}.

The amplitude of a circularly polarized electromagnetic wave transmitted through the stack is given by
\begin{align}\label{eq:Et}
{\cal E}_{\rm tr}^{(s)} =  {\cal E}_{0}^{(s)} + 2\pi \rmi q \sum_n P_n^{(s)} \e^{-\rmi q d n} \,.
\end{align}
By solving Eqs.~\eqref{sys:bi} we calculate ${\cal E}_{\rm tr}^{(s)}$ and the amplitude transmission coefficients for the right-handed and left-handed circularly polarized radiation $t^{(\pm)} = {\cal E}_{\rm tr}^{(\pm 1)}/ {\cal E}_0^{(\pm 1)}$.

Consider now the transmission of linearly polarized light through the chiral stack. The spectrum of transmission is determined by 
\begin{align}
|t(\omega)|^2 = \frac{|t^{(+)}|^2 + |t^{(-)}|^2}{2} \,.
\end{align}
Due to the difference in the transmission coefficients $t^{(+)}$ and $t^{(-)}$, the linearly polarized light, when transmitted through the stack, rotates its polarization plane (optical activity) and acquires ellipticity (circular dichroism). The corresponding rotation angle $\alpha$ and ellipticity angle $\epsilon$ are defined by
\begin{align}\label{alpha-epsilon}
\alpha(\omega) = \frac12\, {\rm arg\,}\frac{t^{(-)} }{ t^{(+)}} \,,\;  
\epsilon(\omega) = {\rm arctan\,}\frac{|t^{(+)}| - |t^{(-)}|}{|t^{(+)}| + |t^{(-)}|} \,.
\end{align}
At $|t^{(+)} - t^{(-)} |\ll |t^{(\pm)}|$, the rotation and ellipticity angles are small and 
Eq.~\eqref{alpha-epsilon} can be rewritten in the compact form 
\begin{align}\label{def:ae}
\alpha(\omega)+ \rmi \epsilon(\omega)= \rmi \, \frac{t^{(+)} - t^{(-)}}{t^{(+)} + t^{(-)}} \,.
\end{align}
Below, we use these relations to analyze optical activity and circular dichroism in bilayers and multi-layer stacks. 

\section{Results and discussion}\label{Sec3}

\subsection{Chiral bilayer}

In a bilayer, the interlayer hopping of excitons leads to the formation of the symmetric and antisymmetric exciton modes denoted by the numbers $m=1$ and $m=2$, respectively~\footnote{The eigen exciton modes are strictly symmetric or antisymmetric only at $\phi=0$.}. Each of them is two-fold degenerate in the spin index $s$. Diagonalization of the Hamiltonian~\eqref{ham:bi} yields the eigen frequencies of the modes
\begin{equation}
\omega^{(1)} = \omega_x - t \,, \; \omega^{(2)} = \omega_x + t \,
\end{equation}
and the eigen functions
\begin{align}\label{exciton_functions}
\Phi^{(s,1)} = \frac1{\sqrt 2} \left( 
\begin{array}{c}
1 \\ \e^{-\rmi s \phi}
\end{array}
\right) \,, \;
\Phi^{(s,2)} = \frac1{\sqrt 2} \left( 
\begin{array}{c}
1 \\ -\e^{-\rmi s \phi}
\end{array}
\right) \,.
\end{align}
The symmetric (bright) mode is efficiently coupled to the electromagnetic field while the coupling of the antisymmetric (dark) mode to the field is very weak~\cite{Ivchenko2005,Ivchenko1994}.

Figure~\ref{fig:bilayer}a shows the transmission spectra of linearly polarized light for the chiral bilayers. Red and blue curves 
correspond to the cases of low and high non-radiative exciton decay rate $\Gamma$. The frequencies of the symmetric and antisymmetric exciton modes are indicated by vertical lines. The transmission spectra feature a single dip at the frequency of the symmetric mode with a barely noticeable trace of the antisymmetric mode.

At $t \gg \Gamma_0$, the symmetric and antisymmetric modes are well separated in frequency and do not interact with each other. It follows from Eq.~\eqref{sys:bi} that the interaction of an individual exciton mode $(s,m)$ with the circularly polarized light propagating along the $z$ axis ($\rightarrow$) or in the opposite direction ($\leftarrow$) is characterized by the rates
\begin{align}\label{def:gamma}
\gamma_{\rightleftarrows}^{(s,m)} = \frac{\Gamma_0}2 \left| \sum_n \Phi_n^{(s,m)*} \e^{\pm \rmi qd n} \right|^2 ,
\end{align}
where $\Phi_n^{(s,m)}$ are the components of the columns~\eqref{exciton_functions}. In particular, the rates of exciton-photon interaction in a bilayer are given by
\begin{align}\label{gamma:sa}
&\gamma_{\rightarrow}^{(s,1)} = \gamma_{\leftarrow}^{(-s,1)} = \frac{\Gamma_0}2 [1 + \cos (s\phi + qd)] \,,\\
&\gamma_{\rightarrow}^{(s,2)} = \gamma_{\leftarrow}^{(-s,2)} = \frac{\Gamma_0}2 [1 - \cos (s\phi + qd)] \,. \nonumber
\end{align}
The total radiative decay rate of excitons in the mode ($s,m$) is $\gamma^{(s,m)} = \gamma_{\rightarrow}^{(s,m)} + \gamma_{\leftarrow}^{(s,m)}$. The symmetric exciton mode is supperradiant and is coupled to light twice stronger than the exciton mode in a single layer. The width of the dip in the transmission spectrum (Fig.~\ref{fig:bilayer}a) associated with the symmetric exciton mode is about $2\Gamma_0 + \Gamma$~\cite{Ivchenko1994,Poshakinskiy2012}. In contrast, the coupling of the antisymmetric mode to light is very weak being determined by the small parameters $qd$ and $\phi$. 

\begin{figure*}[t]
\includegraphics[width=.99\textwidth]{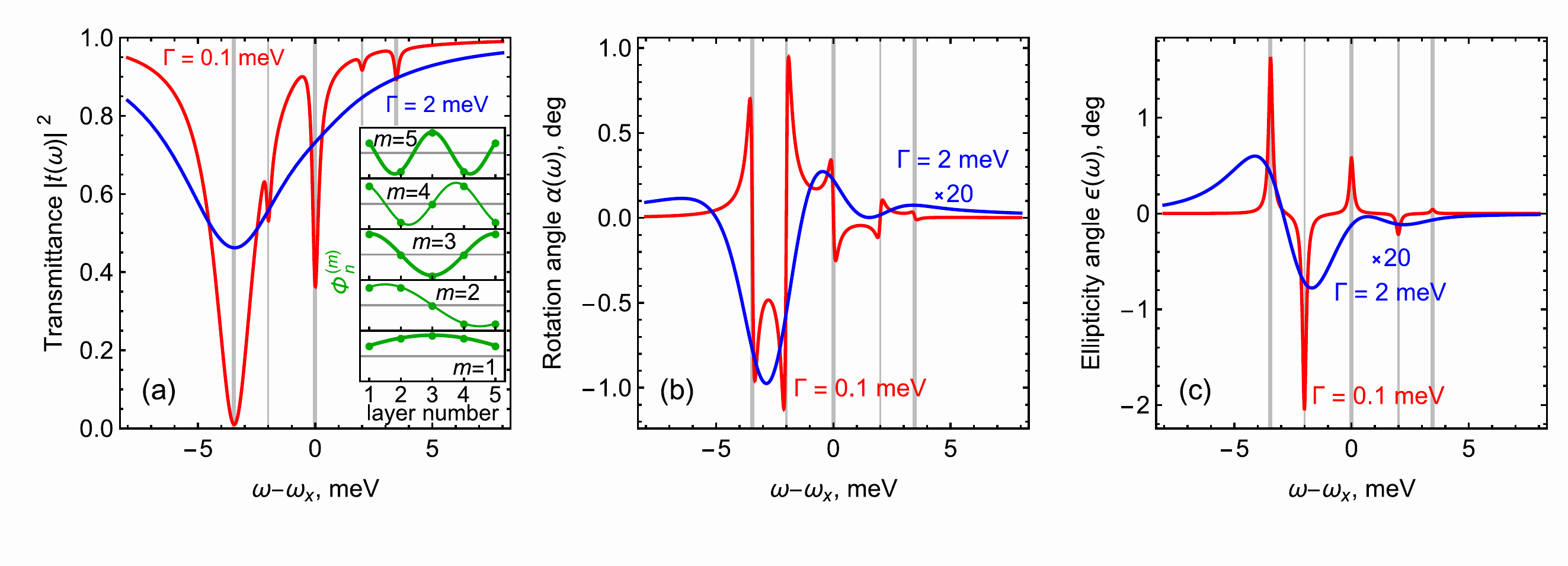}
\caption{(a) Transmittance $|t(\omega)|^2$, (b) polarization rotation angle $\alpha(\omega)$, and (c) ellipticity angle $\epsilon(\omega)$ for a chiral stack of $N=5$ layers. Vertical lines indicate the frequencies of eigen exciton modes. Inset in Fig.~\ref{fig:N5}a shows the distributions of the exciton polarization over the layers for the eigen modes. Red and blue curves are calculated for the stacks with low and high non-radiative decay rates of excitons, respectively. Parameters are the same as for Fig.~\ref{fig:bilayer}.}\label{fig:N5}
\end{figure*}

Figures~\ref{fig:bilayer}b and~\ref{fig:bilayer}c show the frequency dependence of the rotation angle $\alpha(\omega)$ and the 
ellipticity angle $\epsilon(\omega)$, respectively. In contrast to the transmission spectra (Fig.~\ref{fig:bilayer}a),
the spectra of optical rotation (Fig.~\ref{fig:bilayer}b) and circular dichroism (Fig.~\ref{fig:bilayer}c) comprise two resonances of the same strength at the frequencies of both eigen exciton modes. Polarization conversion at the frequency of the dark mode is as effective as that at the frequency the bright mode, whereas the bilayer transparency at the dark mode frequency is very high.
      
This striking result can be explained as follows. The efficiency of polarization conversion at the exciton mode $m$
is determined by the difference of the strengths of the exciton coupling to the right-handed and left-handed circularly polarized light
propagating in the same direction
\begin{equation}
\Delta \gamma_{\rightarrow}^{(m)} = \gamma_{\rightarrow}^{(+1,m)} - \gamma_{\rightarrow}^{(-1,m)} \,.
\end{equation}
These values for the bright and dark exciton modes in the chiral bilayer have the form
\begin{align}
\Delta\gamma_{\rightarrow}^{(1)} = - \Delta\gamma_{\rightarrow}^{(2)} = -\Gamma_0 qd \sin \phi \,,
\end{align}
which follows from Eq.~\eqref{gamma:sa}.
As a result, the bright and dark exciton modes are revealed in the spectra of optical rotation and circular dichroism 
as the resonances of equal strengths and opposite signs.

Analytical solution of Eq.~\eqref{sys:bi} shows that the rotation and ellipticity angles 
in the relevant case of $qd \ll 1$ are given by
\begin{equation}\label{chiral_eqae}
\alpha(\omega) + \rmi \epsilon(\omega) =  \frac{2t \, \Gamma_0 \, qd \, \sin \phi}{(\omega-\omega_x-t+\rmi\Gamma)(\omega-\omega_x+t+\rmi\Gamma)} .
\end{equation}
%
The widths of the resonance features in $\alpha(\omega)$ and $\epsilon(\omega)$ are determined by the nonradiative decay rate $\Gamma$. 
In high-quality structures with $\Gamma \ll \Gamma_0$, the resonances are very sharp and the rotation and ellipticity angles reach
$|\alpha|_{\rm max}, |\epsilon|_{\rm max} \sim (\Gamma_0/\Gamma) qd\, |\sin \phi|$. In the case of $\Gamma \gg t$, the resonances overlap and partly compensate each other leading to the decrease of the rotation angle and the ellipticity angle by the factor $t/\Gamma$.


\subsection{Multi-layer stacks}

The stack consisting of $N$ layers supports $N$ spatial exciton modes. Diagonalization of the Hamiltonian~\eqref{ham:N} yields 
the eigen frequencies of the modes
\begin{align}
\omega^{(m)} = \omega_x - 2t \cos  \frac{ m  \pi}{(N+1)} 
\end{align}
and the eigen functions $\Phi^{(s,m)}$ with the elements
\begin{align}\label{eq:Pn}
&\Phi_n^{(s,m)} = \sqrt{\frac2{N+1}}\sin \frac{n m \pi}{(N+1)} \, \e^{-\rmi n s \phi } \,,
\end{align}
where the index $m$ enumerates the spatial modes and the index $s$ stands for the spin projection. Neglecting the phase factor,
the functions $\Phi^{(s,m)}$ are either even or odd with respect to the center of the stack. The eigen exciton modes in the stack of 
$N=5$ layers are sketched in the inset of Fig.~\ref{fig:N5}a.

The modes are coupled to the electromagnetic field differently. The rates of exciton-photon interaction for the modes $(s,m)$ 
calculated after Eq.~\eqref{def:gamma} have the form
\begin{align}\label{eq:gr}
\gamma_{\rightarrow}^{(s,m)} =  \frac{\Gamma_0\sin^2 \frac{m\pi}{N+1}\{1-(-1)^m \cos [(N+1)(s\phi+qd)]\}}{2(N+1)[\cos \frac{m\pi}{N+1} -\cos(s\phi+qd)]^2} \,
\end{align}
and $\gamma^{(s,m)}_{\leftarrow}=\gamma^{(-s,m)}_{\rightarrow}$. It follows that the modes with odd $m$ are bright and, at $qd, \phi \ll 1$, are characterized by the radiative decay rates 
\begin{align}\label{gamma_odd_m}
\gamma^{(s, m)} =  \frac{2\Gamma_0 \cot^2 \frac{m\pi}{2(N+1)}}{N+1} \qquad \text{(odd $m$)} .
\end{align}
The radiative decay rates of excitons in the modes with even $m$ is much lower,
\begin{align}\label{gamma_even_m}
\gamma^{(s,m)} =  \frac{\Gamma_0(N+1) \cot^2 \frac{m\pi}{2(N+1)}}{2}  [\phi^2+(qd)^2] \;\;\text{(even $m$)} .
\end{align}

In the case of non-interacting modes, $\omega^{(m+1)}- \omega^{(m)} \gg \gamma^{(s,m)} +\gamma^{(s,m+1)}$, the amplitude
transmission coefficient of circularly polarized radiation through the stack is the sum of resonant contributions stemming from individual exciton modes 
\begin{align}
t^{(s)}(\omega) = 1 - \sum_{m=1}^{N} \frac{2 \rmi \gamma^{(s,m)}_{\rightarrow}}{\omega-\omega^{(m)} + \rmi [\gamma^{(s,m)}+\Gamma]}  \,.
\end{align}
Using the definition Eq.~\eqref{def:ae} we obtain the expression for the rotation and ellipticity angles
\begin{align}\label{eq:aesum}
\alpha(\omega) + \rmi \epsilon(\omega) = \sum_{m=1}^N \frac{\Delta\gamma_{\rightarrow}^{(m)}}{\omega-\omega^{(m)} + \rmi\Gamma} \,,
\end{align}
where 
\begin{align}\label{eq:dgamma}
\Delta\gamma_{\rightarrow}^{(m)} = & \Gamma_0\, qd\, \phi \, \frac{\cot^2 \frac{m\pi}{2(N+1)}}{N+1} \\
&\times \left\{ \frac{1-(-1)^m}{ \sin^2 \tfrac{m\pi}{2(N+1)}} + (-1)^m (N+1)^2 ] \right\} \,.\nonumber
\end{align}

Figure~\ref{fig:N5} show the spectra of transmission, optical activity, and circular dichroism of the chiral stacks of $N =5$ layers. Red and blue curves correspond to the stacks with low and high non-radiative exciton decay rates $\Gamma$. The transmission spectrum at low $\Gamma$ (red curve in Fig.~\ref{fig:N5}a) features strong dips at the frequencies of the bright exciton modes (with odd $m$) and weak dips at the frequencies of the dark modes (with even $m$). The strengths of the resonances of both kind decrease with the increase of the mode index $m$. At high $\Gamma$, the resonances associated with individual exciton modes are widen and the fine structure of the transmission spectrum is not resolved. 

The spectral dependences of the rotation angle (Fig.~\ref{fig:N5}b) and the ellipticity angle (Fig.~\ref{fig:N5}c) consist of 5
resonant contributions at the frequencies of eigen exciton modes $\omega^{(m)}$. The resonances are well resolved in the case of low 
non-radiative exciton decay rate $\Gamma$. The strengths of the resonances are determined by $\Delta\gamma_{\rightarrow}^{(m)}$. 
As it is seen in Figs.~\ref{fig:N5}b and ~\ref{fig:N5}c and also follows from Eq.~\eqref{eq:dgamma}, the resonances at the frequencies of bright and dark exciton modes are of comparable strengths and of opposite signs. Interestingly, the strongest resonance occurs at the frequency of the (dark) mode with $m=2$. At high non-radiative decay rate $\Gamma$ (see blue curves in Figs.~\ref{fig:N5}b and ~\ref{fig:N5}c), the resonances corresponding to the neighboring modes overlap and tend to cancel each other since
$\sum_{m=1}^N \Delta\gamma_{\rightarrow}^{(m)} = 0$.

\begin{figure*}[t]
\includegraphics[width=.99\textwidth]{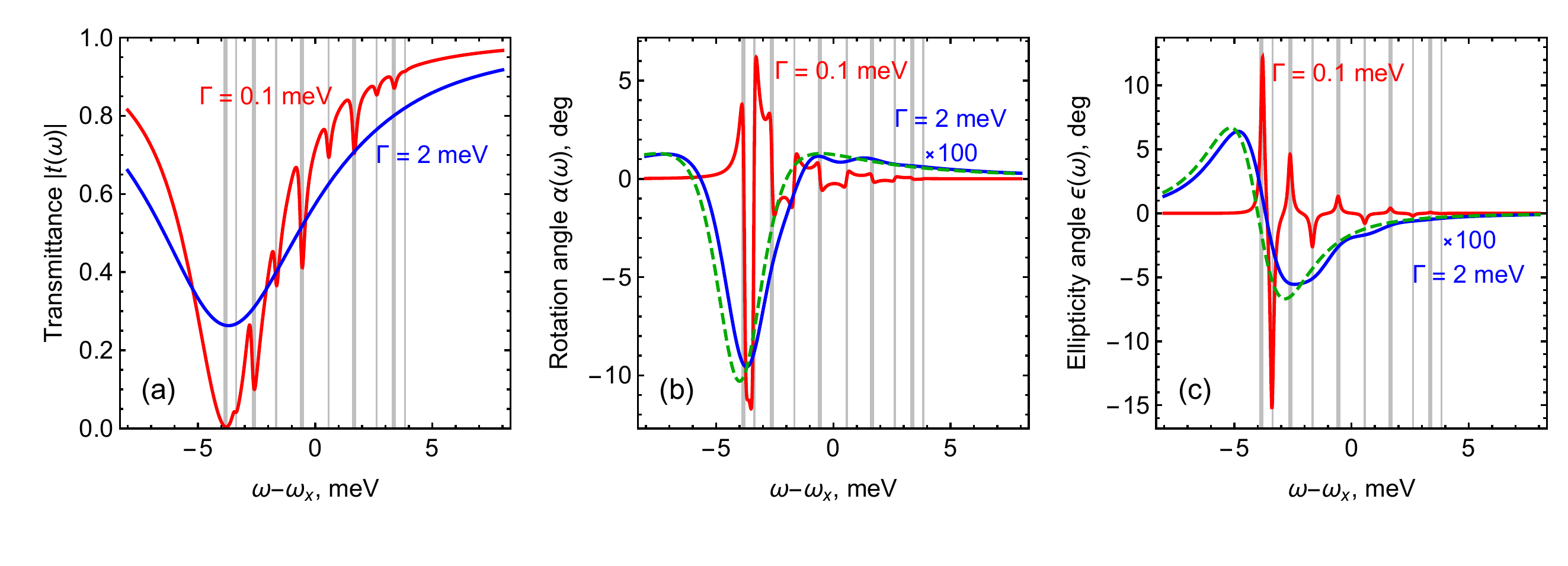} 
\caption{(a) Transmittance $|t(\omega)|^2$, (b) polarization rotation angle $\alpha(\omega)$, and (c) ellipticity angle $\epsilon(\omega)$ for chiral stacks of $N=10$ layers. Vertical lines indicate the frequencies of eigen exciton modes. Red and blue curves are calculated for the stacks with low and high non-radiative decay rates of excitons. Parameters are the same as for Fig.~\ref{fig:bilayer}.
Dashed green curves in Figs.~\ref{fig:N10}b and~\ref{fig:N10}c are plotted after analytical Eq.~\eqref{ae:con} and correspond to the limit of thick stacks or high non-radiative decay rates.}\label{fig:N10}
\end{figure*}

\subsection{Thick stacks}

Figure~\ref{fig:N10} shows the spectra of transmission, optical activity, and circular dichroism for the chiral stack of $N=10$ layers.
In stacks with large $N$, the frequencies of eigen exciton modes, indicated by gray vertical lines in Fig.~\ref{fig:N10}, fill the miniband from $\omega_x -2t$ to $\omega_x +2t$. At large enough $N$, the frequency separation between the neighboring modes $\Delta 
\omega \sim 4 t/N$ becomes smaller than the broadening $\Gamma$ and the individual resonances are overlapped. The resulting angles of
optical rotation and ellipticity are determined by the exciton modes with the frequencies close to the miniband bottom 
$\omega_b = \omega_x -2t$, see blue curves in Fig.~\ref{fig:N10}c and ~\ref{fig:N10}d.

To describe the optical properties of a thick stack, we replace the set of Eqs.~\eqref{sys:bi} for the exciton polarizations 
in individual layers $P_n^{(s)}$ with the differential equation for the continuous function $P^{(s)}(z)$
\begin{align}\label{con:P}
(\omega+\rmi\Gamma)P^{(s)}(z) - \hat H^{(s)} P^{(s)}(z)= - \frac{\Gamma_0}{2\pi q d} \, \mathcal{E}^{(s)}(z)  \,,
\end{align}
where $\hat H^{(s)}$ is the Hamiltonian of excitons with the spin projection $s$ at the miniband bottom
\begin{align}\label{H:cont}
\hat H^{(s)} = \omega_b + \frac{\hat k_z^2}{2M} + 2 t\phi d \, s \hat k_z \,,
\end{align}
$\hat k_z = -\rmi (d/dz)$, $M=1/(2td^2)$ is the effective exciton mass, and $\mathcal{E}^{(s)}(z)$ is the amplitude of the
electric field. The last term in the Hamiltonian~\eqref{H:cont} has the form of spin-orbit interaction linear 
in the exciton wave vector~\cite{Shahnazaryan2015}. It is the term originating from the chirality of the stack that gives rise to the optical 
activity. We also note that the continuous description above is valid provided $qd \,\Gamma_0 \ll \Gamma$, which is well 
fulfilled in realistic structures. 

Equation~\eqref{con:P} together with the Maxwell equation
\begin{align}\label{con:E}
\frac{d^2\mathcal E^{(s)}(z)}{dz^2} + q^2 \mathcal E^{(s)}(z) = - 4\pi q^2   P^{(s)}(z) 
\end{align}
form the closed set of differential equations for the functions $P^{(s)}(z)$ and $\mathcal{E}^{(s)}(z)$. These coupled equations describe
the exciton-polaritons in the stack~\cite{Agranovich_Book,Ivchenko2005}.

The solution of Eqs.~\eqref{con:P} and~\eqref{con:E} in the bulk of the stack has the form $P^{(s)},\mathcal{E}^{(s)}\propto \exp(\rmi Q z)$ with the dispersion $\omega(Q)$ given by
\begin{align}\label{disp}
Q^2(\omega) = q^2 \left( 1- \frac{2 \Gamma_0/ (qd)}{\omega-\omega_b +\rmi\Gamma - Q^2/2M - 2t Qd s\phi }\right) ,
\end{align}
%
where $Q$ is the exciton-polariton wave vector.

Neglecting the mass term in the exciton dispersion and considering the $Q$-linear spin-dependent term as a small correction
we obtain
\begin{equation}\label{polariton_dispersion}
Q(\omega) = \pm Q_0(\omega)  + s \, \delta Q(\omega) \,,
\end{equation}
where
\begin{align}
&Q_0(\omega) = q \sqrt{1-\frac{2\Gamma_0 /(qd)}{\omega-\omega_b +\rmi\Gamma}}  \,,\\
&\delta Q(\omega)= - \frac{2t \Gamma_0 \, q\, \phi}{(\omega-\omega_b +\rmi\Gamma)^2} \,.
\end{align}
\begin{figure}[b]
\includegraphics[width=.66\columnwidth]{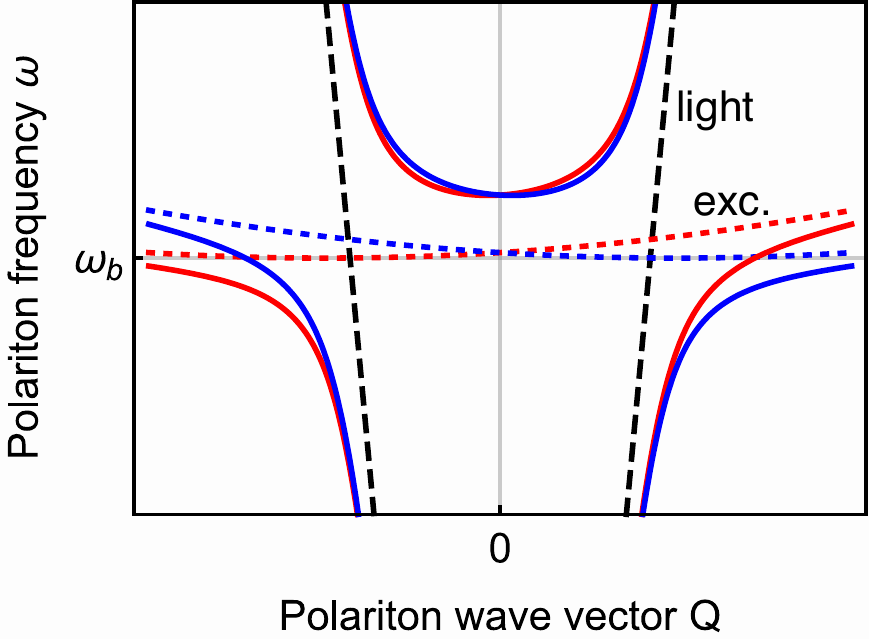} 
\caption{Dispersion of exciton-polaritons in a thick lossless chiral stack (solid red and blue curves). Dashed black curves show the light dispersion $\omega = \pm c Q$. Dashed red and blue curves show the exciton dispersion with included spin-orbit splitting $\omega = \omega_b + Q^2/(2M) \pm 2 t\phi Q d$.}\label{fig:disp}
\end{figure}

Figure~\ref{fig:disp} sketches the dispersions of exciton-polaritons with the spin projections $s=\pm 1$ along the wave vector 
(solid red and blue curves, respectively) given by Eq.~\eqref{disp} for a lossless structure. The exciton-polariton dispersion is
formed as a result of the avoided crossing of the light dispersion (dashed black line) and the exciton dispersion (dashed red and blue curves). The exciton spin-orbit interaction described by the last term in the Hamiltonian~\eqref{H:cont} leads to the spin-orbit splitting of the polariton branches. The splitting is particularly strong for the lower polariton branch at high wave vectors, i.e.,
at the frequencies close to $\omega_b$, where the original exciton splitting is strong and the exciton contribution to the polariton state
is high. The spin-orbit splitting of polariton branches leads to a difference in the transmission coefficients for the right-handed and left-handed radiation.

To calculate the transmission coefficients we solve Eqs.~\eqref{con:P} and~\eqref{con:E} with the polariton dispersion~\eqref{polariton_dispersion} and the boundary conditions of the continuity of the functions $\mathcal{E}^{(s)}$ and $d\mathcal E^{(s)}/dz$ at the front and back surfaces of the stack. Since the major contribution to the rotation and ellipticity angles is proportional to the stack thickness,
small possible spin-dependent corrections to the boundary conditions can be neglected. The calculation yields 
\begin{align}
t^{(s)}(\omega)= \frac{4 Q_0 q  \, \e^{\rmi (s \delta Q - q)L}}
{(Q_0 + q)^2 \e^{-\rmi  Q_0 L} - (Q_0-q)^2 \e^{\rmi  Q_0 L}} \,,
\end{align}
where $L=(N-1)d$ is the stack thickness.


Finally, for the rotation and ellipticity angles in thick chiral stacks we obtain
\begin{eqnarray}\label{ae:con}
\alpha(\omega) + \rmi \epsilon(\omega)=  \frac{2t\, \Gamma_0 \, q L \,  \phi}{(\omega - \omega_b + \rmi\Gamma)^2} \,.
\end{eqnarray}

The rotation and ellipticity angles grow linearly with the stack thickness. The most pronounced conversion of the light polarization occurs for the light frequencies close to $\omega_b$, where the polariton spin-orbit splitting is strongest, see Fig.~\ref{fig:disp}.
The dependences $\alpha(\omega)$ and $\epsilon(\omega)$ calculated after Eq.~\eqref{ae:con} for the stack of 10 layers are shown in 
Figs.~\ref{fig:N10}b and~\ref{fig:N10}c, respectively, by green dashed curves. One can see that the analytical dependences agree well
with the results of exact numerical calculations (blue curves). With the further increase of the number of layers $N$ or the non-radiative decay rate of excitons $\Gamma$, the agreement becomes even better. An estimation after~\eqref{ae:con} yields
$\alpha /L \sim 10^2$~rad/$\mu m$ and $\alpha /L \sim 0.3$~rad/$\mu m$ for stacks with low non-radiative decay rate $\Gamma=0.1$\,meV
and high non-radiative decay rate $\Gamma=2$\,meV, respectively, and the other parameters $\Gamma_0=0.3$\,meV~\cite{Robert2016b}, $t=2$\,meV, $qd=0.01$,
and $\phi=0.1$. 

\section{Polarized photoluminescence of chiral stacks}\label{Sec4}

\begin{figure*}[t]
\includegraphics[width=.99\textwidth]{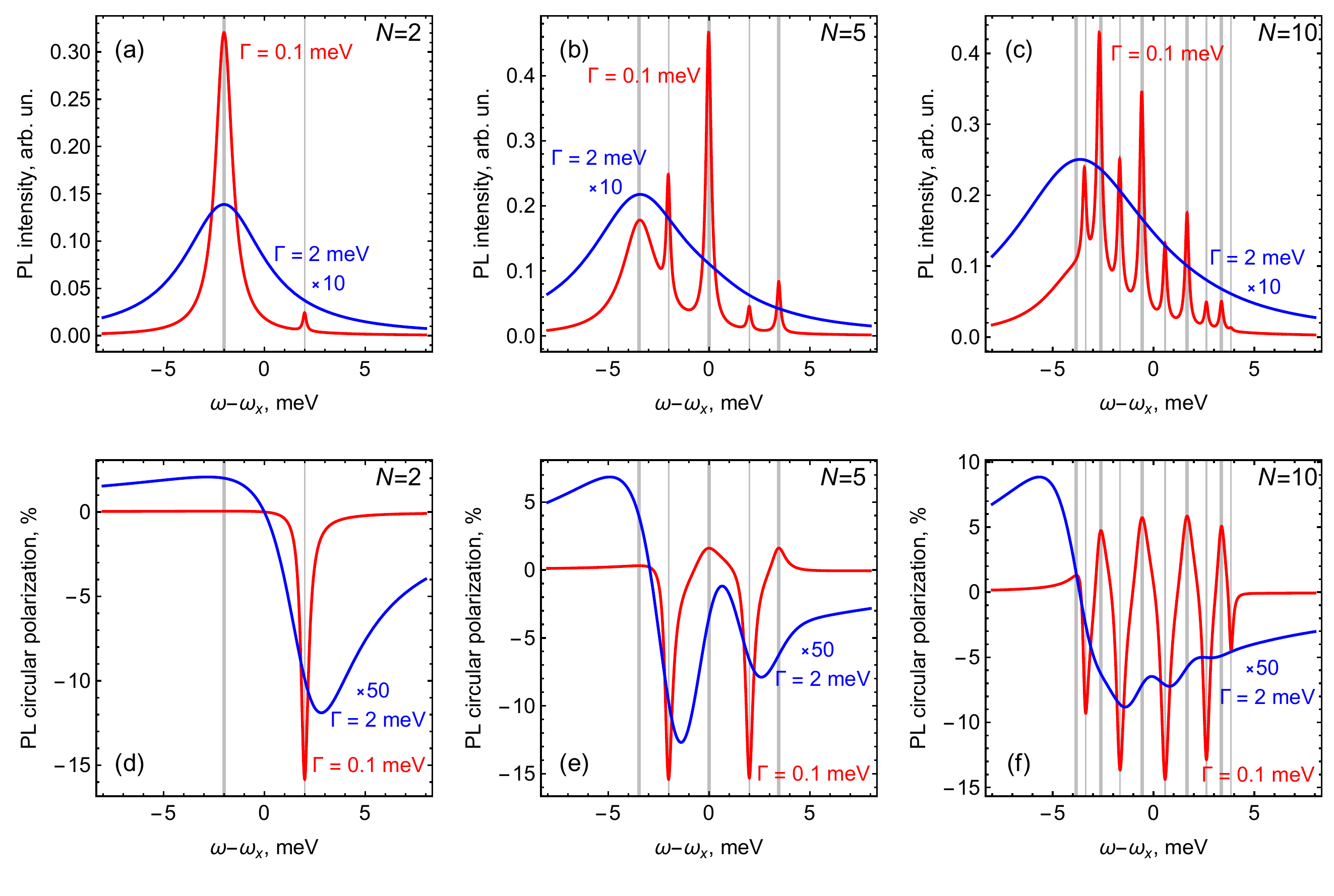} 
\caption{(a)-(c) Photoluminescence spectra and (d)-(f) spectral dependence of the photoluminescence circular polarization degree $P_c$ for the chiral stacks of $N=2$, 5, and $10$ monolayers. Parameters are the same as for Fig.~\ref{fig:bilayer}.}\label{fig:PL}
\end{figure*}

Finally, we discuss the photoluminescence (PL) of chiral stacks and show that the PL is circularly polarized. To calculate the PL spectrum, we apply the approach developed in Ref.~\onlinecite{Voronov2007} for multiple quantum well structures. This approach suggests that the PL is caused by random sources of excitons in the layers. The sources of exciton polarization in the layers are supposed to be identical and incoherent.

In this model, the exciton polarizations in the layers $P_n^{(s)}$ are described by Eq.~\eqref{sys:bi} where the right-hand side is replaced with the rate of polarization generation in the layer $S_n^{(s)}$,
\begin{align}\label{random_sources}
(\omega + \rmi\Gamma) P_n^{(s)} - \sum_{n'} (H_{n,n'}^{(s)} - \rmi \Gamma_0 \e^{\rmi q d|n-{n'}|} ) P_{n'}^{(s)} = S_n^{(s)} \,.
\end{align}
The exciton generations in the layers are independent, therefore, the correlation function has the form $\langle S_n^{(s)} S_{n'}^{(s')} \rangle = S^2(\omega) \delta_{n,n'} \delta_{s,s'}$.

The amplitude of the electric field of the radiation emitted by the stack in the $+z$ direction is given by Eq.~\eqref{eq:Et} with ${\cal E}_0^{(s)}=0$. This yields the expression for the spectral density of the emitted radiation
\begin{align}\label{eq:Isum}
I_{\rightarrow}^{(s)}(\omega) \propto q^2 \left|  \sum_{n} P^{(s)}_{n} \e^{-\rmi q d n} \right|^2 \,
\end{align}
and $I_{\leftarrow}^{(s)}(\omega)=I_{\rightarrow}^{(-s)}(\omega)$. 

Following the procedure described in the previous section we calculate the exciton polarizations~\eqref{random_sources} in the layers and then the PL spectrum~\eqref{eq:Isum}. Particularly, in the case of non-interacting exciton modes in the multi-layer structure, the spectral density of the PL with the certain circular polarization $s$ assumes the form
\begin{align}\label{Irsum}
I_{\rightarrow}^{(s)}(\omega) \propto q^2 S^2 \sum_m \frac{2\gamma_{\rightarrow}^{(s,m)}}{(\omega-\omega^{(m)})^2 + (\gamma^{(m)}+\Gamma)^2} \,.
\end{align}

Figures~\ref{fig:PL}a-\ref{fig:PL}c show the total PL spectra $I_{\rightarrow} = I_{\rightarrow}^{(+1)} + I_{\rightarrow}^{(-1)}$ of chiral stacks consisting of $N=2$, 5, and 10 monolayers. The spectra are calculated following Eqs.~\eqref{random_sources} and~\eqref{eq:Isum} for the frequency-independent $S^2$ which corresponds to the pumping of different exciton states with the equal probability. Similarly to the spectra of transmission (Figs.~\ref{fig:bilayer}a,~\ref{fig:N5}a, and~\ref{fig:N10}a),
the PL spectra in the stacks with low non-radiative decay rate $\Gamma$ comprise $N$ peaks associated with the eigen exciton modes.
The peaks at the frequencies of bright exciton modes (indicated by thick vertical lines) are stronger than the peaks
at the frequencies of dark exciton modes (indicated by thin vertical lines). In the stacks with high non-radiative decay rate $\Gamma$ the individual peaks are broadened forming the smooth PL spectra.

Figures~\ref{fig:PL}d-\ref{fig:PL}f show the spectral dependence of the PL circular polarization degree defined by
\begin{align}\label{def:Pc}
P_c (\omega) = \frac{I_{\rightarrow}^{(+)}(\omega) - I_{\rightarrow}^{(-)}(\omega)}{I_{\rightarrow}^{(+)}(\omega) + I_{\rightarrow}^{(-)}(\omega)} \,.
\end{align}
The PL is circularly polarized because of the chiral stacking of the layers. In structures with low non-radiative decay rate $\Gamma$ (red curves), the degree of circular polarization at the frequencies of  dark exciton modes is much higher than that at the frequencies of bright modes. The reason is that $P_c$ of the radiation emitted by the $m$-th exciton mode is determined by the ratio $\Delta\gamma_{\rightarrow}^{(m)}/\gamma^{(m)}$, as follows from Eqs.~\eqref{Irsum} and~\eqref{def:Pc}. While $\Delta\gamma_{\rightarrow}^{(m)}$ is of the same order for bright and dark modes [Eq.~\eqref{eq:dgamma}], much smaller value of $\gamma^{(m)}$ for dark modes than for bright modes [Eqs.~\eqref{gamma_odd_m} and~\eqref{gamma_even_m}] leads to the higher degree of circular polarization. In stacks with high $\Gamma$ (blue curves), the contributions of individual exciton modes overlap and partly compensate each other leading to the decrease of the PL circular polarization degree.

\section{Summary}\label{Sec5}

To summarize, we have developed a microscopic theory of optical activity and circular dichroism in chiral stacks of two-dimensional crystals, such as the layers of transition metal dichalcogenides, in the spectral range of exciton transitions. The theory takes into account spin-dependent transport of excitons between the layers of the stack and coupling of excitons to an electromagnetic field. We have shown that the frequency dependence of the optical rotation angle and the ellipticity angle exhibits complex behavior reflecting the structure of eigen exciton modes in chiral stacks. In stacks with low non-radiative decay rate of excitons, the spectra of optical activity and circular dichroism comprise sharp resonances associated with individual exciton modes. In stacks with high non-radiative decay rate, the individual resonances overlap and partially compensate each other. The spectra of optical activity and circular dichroism in thick chiral stacks is well described by the developed analytical theory of exciton-polaritons with the effective spin-orbit coupling. We have also calculated the spectra of exciton photoluminescence of chiral stacks in the conditions of non-resonant pumping and shown that the photoluminescence is circularly polarized. 

\section*{Acknowledgments}
We acknowledge fruitful discussions with A.N.~Poddubny, E.L.~Ivchenko, and L.E.~Golub.
This work was supported by the Government of the Russian Federation (Project No. 14.W03.31.0011 at the Ioffe Institute). A.V.P. also acknowledges the support from the RF president grant SP-2912.2016.5 and the ``Basis'' Foundation.

%

\end{document}